\documentclass[12pt]{iopart}
\newcommand{\im}[0]{\rm{Im}\,}
\newcommand{\re}[0]{\rm{Re}\,}
\usepackage{graphicx}% Include figure files
\begin{document}

\title[]{Bounds for the refractive indices of metamaterials}
\author{Johannes Skaar}
\address{Department of Electronics and Telecommunications, Norwegian University of Science and Technology, NO-7491 Trondheim, Norway}
\ead{johannes.skaar@iet.ntnu.no}

\author{Kristian Seip}
\address{Department of Mathematical Sciences, Norwegian University of Science and Technology, NO-7491 Trondheim, Norway}
\ead{seip@math.ntnu.no}

\begin{abstract}
The set of realizable refractive indices as a function of frequency is considered. For passive media we give bounds for the refractive index variation in a finite bandwidth. Special attention is given to the loss and index variation in the case of left-handed materials.
\end{abstract}

\maketitle

\section{Introduction}
During recent years several new types of artificial materials or metamaterials with sophisticated electromagnetic properties have been designed. The fabrication of custom structures with dimensions much smaller than the relevant wavelength has made it possible to tailor the effective electric permittivity $\epsilon$ and the magnetic permeability $\mu$. For example, materials with thin metal wires simulate the response of a low-density plasma so that $\re\epsilon$ may become negative in the microwave range \cite{pendry1996}. Similarly, with the help of a split-ring structure, a strong magnetic resonance is achieved so that $\re\mu$ may be negative \cite{pendry1999}. Passive media with $\re\epsilon$ and $\re\mu$ simultaneously negative, first realized by Smith {\it et al.} \cite{smith}, are particularly interesting. Such materials are often referred to as left-handed, since, for negative $\epsilon$ and $\mu$, the electric field, the magnetic field, and the wave vector form a left-handed set of vectors. As the Poynting vector and the wave vector point in opposite directions, the refraction at the boundary to a regular medium is negative. The concept of negative refraction, introduced by Veselago already in 1968 \cite{veselago}, has opened a new horizon of applications in electromagnetics and optics. In particular the possibility of manipulating the near-field may have considerable potential, enabling imaging with no limit on the resolution \cite{pendry2000}.

Materials with negative $\epsilon$ and $\mu$ are necessarily dispersive \cite{landau_lifshitz_edcm,veselago}, and loss is unavoidable. Loss has serious consequences for the performance of certain components; for example, it has been shown that the resolution associated with the Veselago--Pendry lens is strongly dependent on the loss of the material \cite{ramakrishna2002,nieto-vesperinas}. Therefore, it is important to look for metamaterial designs with negative real part of the refractive index while the loss is low. In this paper, instead of performing a search in an infinite, complex design space, we will find ultimate, theoretical bounds based on causality. We will also find optimal $\epsilon(\omega)$ and $\mu(\omega)$ functions. For example, suppose our goal is refractive index close to --1 while loss is negligible in a limited bandwidth $\Omega\equiv[\omega_1,\omega_2]$. What is then the minimal variation of the refractive index in $\Omega$, given that the medium is passive? If we force the real part of the refractive index to be exactly --1 in $\Omega$, what will then be the minimal loss there?

It is common to assume that the medium can be described by some specific model, such as, for example, single or multiple Lorentzian resonances. While this permits a straightforward analysis, it is not clear if a more general medium would give a more optimal response in some sense. We will therefore not use a spesific model, but rather assume only causality.

\section{Realizable electromagnetic parameters}
Any electromagnetic medium must be causal in the microscopic sense; the polarization and magnetization cannot precede the electric and magnetic fields, respectively. This means that $\epsilon(\omega)$ and $\mu(\omega)$ obey the Kramers--Kronig relations. In terms of the susceptibilities $\chi=\epsilon-1$ or $\chi=\mu-1$, these relation can be written
\begin{eqnarray}
\im \chi=\mathcal H\,\re \chi,\label{KKchi1}\\
\re \chi=-\mathcal H\, \im \chi,\label{KKchi2}
\end{eqnarray}
where $\mathcal H$ denotes the Hilbert transform \cite{nussenzveig}. These conditions are equivalent to the fact that $\chi$ is analytic in the upper half-plane ($\im\omega>0$), and uniformly square integrable in the closed upper half-plane \footnote{If the medium is conducting at zero frequency, the electric $\chi$ is singular at $\omega=0$. Although $\chi$ is not square integrable in this case, similar relations as (\ref{KKchi1})-(\ref{KKchi2}) can be derived \cite{landau_lifshitz_edcm}.}. The susceptibilities are defined for negative frequencies by the symmetry relation
\begin{equation}
\chi(-\omega)=\chi^*(\omega), \label{sym}
\end{equation}
so that their inverse Fourier transforms are real. For passive media, in addition to (\ref{KKchi1})-(\ref{sym}) we have: 
\begin{equation}
\im \chi(\omega)>0 \qquad\rm{for}\ \, \omega>0. \label{loss}
\end{equation}
The losses, as given by the imaginary parts of the susceptibilities, can be vanishingly small; however they are always present unless we are considering vacuum \cite{landau_lifshitz_edcm}. 

Eqs. (\ref{KKchi1})-(\ref{loss}) imply that $1+\chi$ is zero-free in the upper half-plane \cite{landau_lifshitz_edcm}. Thus the refractive index $n=\sqrt{\epsilon}\sqrt{\mu}$ can always be chosen as an analytic function in the upper half-plane. With the additional choice that $n\to +1$ as $\omega\to\infty$, $n$ is determined uniquely, and it is easy to see that (\ref{KKchi1})-(\ref{loss}) hold for the substitution $\chi\to n-1$.

While any refractive index with positive imaginary part can be realized at a single frequency, the conditions (\ref{KKchi1})-(\ref{loss}) put serious limitations on what is possible to realize in a finite bandwidth. First we will investigate the possibility of designing materials with the real part of the refractive index less than unity. In particular we will analyze to what extent it is possible in a limited bandwidth to have a constant index below unity (or even below zero) while the loss is small. We set $n-1=u+iv$ (or $\chi=u+iv$), where $u$ and $v$ are the real and imaginary parts of $n-1$ (or $\chi$), respectively. To begin with, we set
$v(\omega)=0$ in the interval $\Omega=[\omega_1,\omega_2]$. (The case with a small imaginary part will be treated later.) By writing out the Hilbert transform and using (\ref{sym}), we find 
\begin{equation}
\label{uint} 
u(\omega)=-\frac{2}{\pi}\int_0^{\omega_1}
\frac{v(\omega')\omega'\rmd\omega'}{\omega^2-\omega'^2}
+\frac{2}{\pi}\int_{\omega_2}^\infty
\frac{v(\omega')\omega'\rmd\omega'}{\omega'^2-\omega^2}
\end{equation}
for $\omega\in\Omega$. Note that both terms in (\ref{uint}) are increasing functions of $\omega$. Since the goal is a constant, negative $u(\omega)$ in $\Omega$, the second term should be as small as possible. In the limit where the second term is zero, we obtain
\begin{equation}
u(\omega)-u(\omega_1)=\frac{2}{\pi}\int_0^{\omega_1}
\frac{v(\omega')\omega'\rmd\omega'}{\omega_1^2-\omega'^2}\frac{\omega^2-\omega_1^2}{\omega^2-\omega'^2},
\end{equation}
and therefore
\begin{equation}
\label{variationbound}
u(\omega)-u(\omega_1)>|u(\omega_1)|\frac{\omega^2-\omega_1^2}{\omega^2}, \quad \omega\in\Omega,
\end{equation}
provided $u(\omega_1)$ is negative. In particular, the largest variation in the interval is
\begin{equation}
\label{variationboundOmega}
u(\omega_2)-u(\omega_1) >|u(\omega_1)|(2\Delta-\Delta^2).
\end{equation}
Here we have defined the normalized bandwidth $\Delta=(\omega_2-\omega_1)/\omega_2$. These bounds are realistic in the sense that equality is obtained asymptotically when $v(\omega)$ approaches a delta function in $\omega=0^+$. In this limit $u(\omega)=u(\omega_1)\omega_1^2/\omega^2$.

It is interesting to estimate how much loss we must allow in the interval to wash out the variation (\ref{variationbound}). Letting $v(\omega)$ approach a delta function in $\omega=0^+$, and adding the positive function $|u(\omega_1)|\sqrt{\omega_2^2-\omega^2}\sqrt{\omega^2-\omega_1^2}/\omega^2$ for $\omega\in\Omega$ correspond to a constant $u(\omega)=u(\omega_1)$ in $\Omega$. Furthermore, it can be shown that this particular $v$ corresponds to the minimal possible loss in the interval. The proof for this claim is given elsewhere \cite{seip2005}. Thus the maximal value of the (minimal) loss in the interval satisfies
\begin{equation}
\label{maxloss}
v_{\max}>|u(\omega_1)|\frac{\omega_2^2-\omega_1^2}{2\omega_1\omega_2}=|u(\omega_1)|\Delta+O(\Delta^2).
\end{equation}

By a superposition of the optimal solutions associated with the bounds (\ref{variationbound}) and (\ref{maxloss}), we obtain a bound for the loss when a certain fraction $1-\alpha$ ($0\leq\alpha\leq 1$) of the variation (\ref{variationbound}) remains:
\begin{equation}
\label{maxlossvar}
v_{\max} > \alpha|u(\omega_1)|\frac{\omega_2^2-\omega_1^2}{2\omega_1\omega_2}.
\end{equation}

As an example, consider the case where the goal is refractive index close to --1 in an interval $\Omega$ with $\Delta\ll 1$. In the limit of zero imaginary index, (\ref{variationboundOmega}) gives that the variation of the real index in the interval is larger than $4\Delta$. It is interesting that the minimal variation is obtained approximately if the medium has sharp Lorentzian resonances for a low frequency. For example, let $\epsilon(\omega)=\mu(\omega)=1+\chi(\omega)$, where
\begin{equation}
\label{lorentz}
\chi(\omega)=\frac{F\omega_{0}^2}{\omega_{0}^2-\omega^2-i\omega\Gamma}.
\end{equation}
Here, $F$, $\omega_{0}$, and $\Gamma$ are positive parameters. If the bandwidth $\Gamma$ and center frequency $\omega_{0}$ are much smaller than $\omega_1$, $\re n(\omega)\approx 1-F(\omega_{0}/\omega)^2$ and $\im n(\omega)\approx F\omega_{0}^2\Gamma/\omega^3$ for $\omega\geq\omega_1$. If we require $\re n(\omega_1)=-1$, we obtain $\im n(\omega_1)\approx 2\Gamma/\omega_1$ and $\re n(\omega_2)-\re n(\omega_1)\approx 4\Delta$. When $\Gamma/\omega_1\to 0$, this corresponds to the optimal refractive index function
associated with the bound (\ref{variationboundOmega}). Furthermore, it is worth noting that if we want the real index variation to be zero in $\Omega$, the maximum imaginary part of the refractive index in $\Omega$ must be larger than $2\Delta$. The required imaginary part in $\Omega$ can roughly be approximated by weak resonances at $(\omega_1+\omega_2)/2$, see Fig. 1.
\begin{figure}%[t!]
\begin{center}
\includegraphics[height=6.7cm,width=8.1cm]{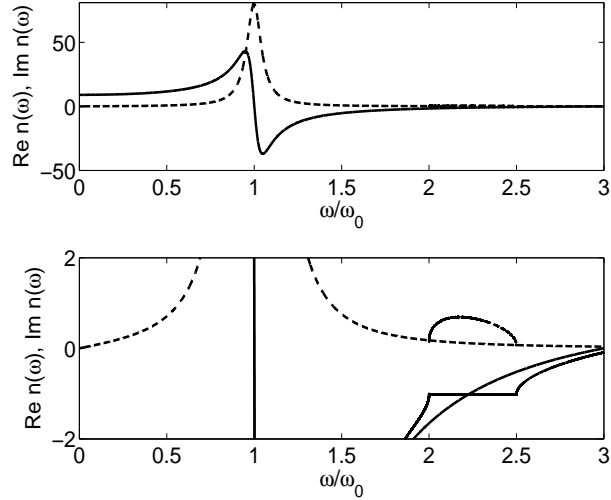}
\end{center}
\caption{The real (solid line) and imaginary (dashed line) refractive index associated with a Lorentzian resonance at $\omega=\omega_{0}$. The two figures represent the same functions but the scale is different. Also shown are the real part and imaginary part for the case where the real part is --1 in $\Omega$. In general this refractive index function can be found using the approach in \cite{seip2005}; however, when $\omega_1$ is much larger than the resonance frequency and bandwidth, the required $v$ in $\Omega$ is $2\sqrt{\omega_2^2-\omega^2}\sqrt{\omega^2-\omega_1^2}/\omega^2$. The parameters used are $\omega_1=2\omega_{0}$, $\omega_2=2.5\omega_{0}$, $\Gamma=0.1\omega_{0}$, and $F=8$.}
\label{fig:resonance}
\end{figure}

So far we have considered  the case where the goal is a constant $u(\omega)<0$ in $\Omega$. If the goal is $u(\omega)>0$ in $\Omega$, it is the last term in (\ref{uint}) that comes to rescue. Inspired by the result (\ref{variationbound}), we may let $u(\omega)$ approach a delta function at a frequency much larger than $\omega$. Indeed, in the limit where this resonance frequency approaches infinity, the function $u(\omega)$ is constant and positive in $\Omega$ while $v(\omega)$ is zero. Of course this limit is not realistic; in practice the resonance frequency is limited to, say $\omega_{\max}$, where $\omega_{\max}>\omega_2$. The associated bounds are easily deduced along the same lines as above. For example, (\ref{variationboundOmega}) becomes
\begin{equation}
\label{variationboundOmegaomegamax}
u(\omega_2)-u(\omega_1) >u(\omega_1)\frac{\omega_2^2-\omega_1^2}{\omega_{\max}^2-\omega_2^2}.
\end{equation}
Similarly, there may be a lower bound $\omega_{\min}$, where $0<\omega_{\min}<\omega_1$, on the resonance frequency. The stricter inequalities in this case, corresponding to (\ref{variationbound})-(\ref{maxlossvar}), can be found in a similar fashion.

Eq. (\ref{variationboundOmega}) and (\ref{variationboundOmegaomegamax}) have another interesting consequence. If the loss is zero in an infinitesimal bandwidth around $\omega$, we immediately find that the derivative $\rmd u/\rmd\omega$ is bounded from below:
\begin{equation}
\label{dnbound}
\frac{\rmd u}{\rmd\omega} >
\cases{2|u(\omega)|/\omega & for $u(\omega)<0$, \\
  0 & for $u(\omega)\geq 0$.}
\end{equation}
For the case $u(\omega)\geq 0$ we have set $\omega_{\max}=\infty$. Note that also this bound is tight. A similar bound was obtained previously for $\epsilon(\omega)$ and $\mu(\omega)$ \cite{landau_lifshitz_edcm}. Eqs. (\ref{dnbound}) should also be compared to the weaker bound $\rmd n/\rmd\omega>|u(\omega)|/\omega$ which was obtained recently \cite{smith_kroll}. While the latter bound means that the group velocities of transparent, passive media are bounded by $c$, (\ref{dnbound}) implies the maximum group velocity $c/(2-n)$ for $n<1$ (and trivially $c/n$ for $n \geq 1$). Here $c$ is the vacuum light velocity. 

When the loss in a bandwidth $\Omega$ is at most $v_{\max}$, (\ref{dnbound}) becomes
\begin{equation}
\label{dnboundloss}
\frac{\rmd u}{\rmd\omega} >
\cases{\left(\frac{2|u(\omega)|}{\omega}-\frac{4v_{\max}}{\pi\omega\Delta}\right) & for $u(\omega)<0$, \\
 -\frac{4v_{\max}}{\pi\omega\Delta} & for $u(\omega)\geq 0$,}
\end{equation}
to lowest order in $\Delta$, for $\omega$ close to $(\omega_1+\omega_2)/2$. In obtaining (\ref{dnboundloss}) we have assumed that $v(\omega)$ is approximately constant in $\Omega$, and calculated the corresponding contribution to $\rmd u/\rmd\omega$. A similar bound can be derived when $v(\omega)$ varies slowly in $\Omega$; for example, if $v(\omega)$ is the imaginary part of a Lorentzian with $\Gamma=\omega_2-\omega_1$, the inequality holds with the replacement $4/\pi\to 2$. Note that without an assumption on the variation of $v(\omega)$, $\rmd u/\rmd\omega$ can take any value.

A similar method as that leading to (\ref{variationboundOmega}) can be used to find bounds for the variation of derivatives in $\Omega$, in the limit of no loss. For the first order derivative, the variation can be arbitrarily small to first order in $\Delta$, for any positive $\rmd u(\omega_1)/\rmd\omega$. For negative second order derivative $D\equiv \rmd^2 u/\rmd\omega^2$ (first order dispersion coefficient) we obtain
\begin{equation}
\label{DvariationboundOmega}
D(\omega_2)-D(\omega_1) > |D(\omega_1)|4\Delta+O(\Delta^2).
\end{equation}

\section{Discussion and conclusion}
We have considered the set of realizable permittivities, permeabilities and refractive indices. For passive media we have used (\ref{KKchi1})-(\ref{loss}) to prove ultimate bounds for the loss and variation of the real part of the permittivity, permeability, and refractive index. 

While the notation has indicated an isotropic medium, the bounds in this paper are valid for the effective index of the normal modes of anisotropic media as well. In this case the identification of associated $\epsilon$ and $\mu$ tensors from the effective index is more complicated, but nevertheless feasible. More generally, the bounds are valid for the effective index of any electromagnetic mode that can be excited separately and causally, provided the effective index is independent of the longitudinal coordinate.

On the basis of causality, it is clear that the susceptibilities of active media also satisfy (\ref{KKchi1})-(\ref{sym}). However, (\ref{loss}) is certainly not valid. Kre{\u\i}n and Nudel'man have shown how to approximate a square integrable function in a finite bandwidth by a function satisfying (\ref{KKchi1})-(\ref{sym}) \cite{nudelman,krein2}. The approximation can be done with arbitrary precision; however, there is generally a trade-off between precision and the energy of $\chi$ outside the interval \cite{skaar2003}. Once a valid susceptibility has been found, a possible refractive index can be found e.g. by setting $n=\epsilon=\mu=1+\chi$. Hence, in principle, for active media $n$ can approximate any square integrable function in a limited bandwidth.

\section*{References}
%\bibliographystyle{unsrt}
%\bibliography{causbib}% Produces the bibliography via BibTeX.

\end{document}